\documentclass[prb,showpacs,twocolumn,aps,superscriptaddress,a4paper]{revtex4-1}
\usepackage{dcolumn,amssymb,amsmath,amsfonts,graphicx,latexsym,color,braket}
\usepackage{notes2bib}
\usepackage{multirow}

\definecolor{myblue}{rgb}{0,0,0.75}

\begin{document}

\title{Discrete Lorentz symmetry and discrete spacetime translational
symmetry in two- and three-dimensional crystals}

\author{Xiuwen Li}
\affiliation{Department of Physics, Zhejiang Normal University, Jinhua 321004, China}

\author{Jiaxue Chai}
\affiliation{Department of Physics, Zhejiang Normal University, Jinhua 321004, China}

\author{Huixian Zhu}
\affiliation{Department of Physics, Zhejiang Normal University, Jinhua 321004, China}

\author{Pei Wang}
\email{wangpei@zjnu.cn}
\affiliation{Department of Physics, Zhejiang Normal University, Jinhua 321004, China}

\date{\today}

\begin{abstract}
As is well known, crystals have discrete space translational symmetry.
It was recently noticed that one-dimensional crystals possibly have discrete Poincar\'{e} symmetry,
which contains discrete Lorentz and discrete time translational symmetry as well.
In this paper, we classify the discrete Poincar\'{e} groups on two- and three-
dimensional Bravais lattices. They are the candidate symmetry groups of two- or three-dimensional
crystals, respectively. The group is determined by an integer generator $g$, and it
reduces to the space group of crystals at $g=2$.
\end{abstract}

\maketitle

\section{Introduction}

After the papers by Wilczek and Shapere~\cite{Shapere:2012gq,Wilczek:2012jq}, the spontaneous breaking
of time translational symmetry (TTS) becomes a focus of research and
controversy~\cite{Bruno:2013gt,Watanabe:2015jh,Else:2016gf,Sacha:2015hb,Khemani:2016gd,Zhang:2017,Choi:2017}.
According to the special theory of relativity, time and space must be put on the same footing.
The existence of crystals is a manifestation of space translational symmetry (STS)
being spontaneously broken into a discrete one~\cite{Marder}.
It is then natural to think that TTS can also be spontaneously broken into a discrete one.

It was soon noticed that the original definition of TTS breaking
by Wilczek is problematic. Bruno~\cite{Bruno:2013gt}, and
Watanabe and Oshikawa~\cite{Watanabe:2015jh}, proved
that continuous TTS cannot be broken in the ground state or Gibbs ensemble
of a quantum system. But this does not rule out the possibility of TTS breaking in nonequilibrium states.
The spontaneous breaking of TTS was redefined for
periodically driven systems~\cite{Else:2016gf}. The proposals for experiments were
discussed~\cite{Sacha:2015hb,Khemani:2016gd} and realized in 2016~\cite{Zhang:2017,Choi:2017}.

In the theory of special relativity, STS and TTS
are connected to each other by a rotation in spacetime, i.e. the Lorentz transformation.
Lagrangian must be invariant under both the Lorentz transformations
and spacetime translations, which combine into the Poincar\'{e} group~\cite{Weinberg}.
The low-energy states may have less symmetries.
After a process called spontaneous symmetry breaking, the symmetry group reduces to
a subgroup of the Poincar\'{e} group. Examples are the space groups of crystals.
It was long believed that crystals have discrete STS,
continuous TTS, but no Lorentz symmetry.

An exceptional possibility was discussed in Ref.~[\onlinecite{Wang}].
In 1+1 dimensions, the Poincar\'{e} group has subgroups that
include both discrete spacetime translations and discrete Lorentz transformations.
A crystal cannot have continuous Lorentz symmetry, but it can have discrete one.
As thus, it must have discrete TTS as well. The overall translational symmetry
is determined by a lattice in 1+1 dimensions. And the period of TTS is
connected to the lattice constant of crystal and the speed of light.
The possibility of crystals owning discrete TTS and Lorentz symmetry has not been noticed before.
Indeed, no observation of such symmetry was reported up to now. A possible
explanation is that the period of TTS is too small, only in the order of $10^{-18}s $.

In this paper, we generalize above results to higher dimensions.
We classify the discrete Poincar\'{e} groups in 1+2 and 1+3 dimensions. They are the
candidate symmetry groups of the corresponding two- and three-dimensional crystals, respectively.
The STS of a crystal is given by its Bravais lattice. There are 5 and 14
Bravais lattices in two and three spatial dimensions, respectively.
We find discrete Poincar\'{e} symmetry only on six of them.
Table~\ref{tab:1} enumerates these lattices and the discrete symmetry on them.
Our results have potential application in the search of exotic symmetry of crystals.
Our finding might also be interesting to ones who study quantum gravity.
Some approaches to quantum gravity propose the spacetime to
have a lattice structure~\cite{Sorkin87,Sorkin03,Yamamoto,Livine,Rieffel00,Rieffel01}.
And the study on discrete Poincar\'{e} groups
provides a way of maintaining the Lorentz symmetry on a spacetime lattice.

\begin{table}[h]\label{tab:1}
\renewcommand\arraystretch{1.5}
\begin{tabular}{| c | c | c | c | c | c | c | c |}
\hline
 \multirow{2}{*}{}  & \multirow{2}{*}{1D} & 2D & \multicolumn{5}{c|}{3D} \\ \cline{3-8}
 & & $R_e$ & $M$ & $O$ & $O_b$ & $T$ & $H$  \\
\hline
\textbf{R} & $ I$ & $D_2$ & $C_{2h}$ & $D_{2h}$ & $D_{2h}$ & $D_{4h}$ 
& $D_{6h}$ \\
\hline

\multirow{4}{*}{\textbf{Y}} & \multirow{2}{*}{$n_0 r_0+n_1 r_1$} & $n_0 r_0+n_1 r_1$ &
\multicolumn{5}{c|}{$n_0 r_0+n_1 r_1 + n_2 r_2$ } \\ 
 & & $ + n_2 r_2$ & \multicolumn{5}{c|}{$+n_3 r_3$ }\\
\cline{2-8}

 & \multicolumn{7}{c|}{$r_1=\left( 0, a_1, 0, \cdots \right)^T$, $r_2 \perp r_1$, $r_3 \perp r_1$,} \\ 
& \multicolumn{7}{c|}{
$r_0=\left({a_1\sqrt{g^{2}-4}}/{\left(2c\right)},0,0,\cdots \right)^{T}$ for even $g$,} \\
& \multicolumn{7}{c|}{$r_0=\left({a_1\sqrt{g^{2}-4}}/{\left(2c\right)},a_1/2,0,\cdots \right)^{T}$ for odd $g$.}\\
\hline

\textbf{L} &\multicolumn{7}{c|}{
$\left\{ RB_{\vec{v}_j} \big| R\in \textbf{R}, \ \vec{v}_j \parallel r_1 \right\} $ }\\
\hline

$\mathcal{P}$ & \multicolumn{7}{c|}{$\mathcal{P}=\textbf{L}\times \textbf{Y}$} \\
\hline
\end{tabular}
\caption{Discrete Poincar\'{e} groups (denoted by $\mathcal{P}$)
on the one-dimensional (1D), two-dimensional (2D)
and three-dimensional (3D) Bravais lattices. $R_e$ denotes the rectangular lattice.
$M$, $O$, $O_b$, $T$ and $H$ denote the monoclinic, orthorhombic,
base-centered orthorhombic, tetragonal and hexagonal lattices, respectively.
$\textbf{R}$ is the point group of the corresponding Bravais lattice.
The set $\mathcal{P}$ is a direct product of $\textbf{L}$ and $\textbf{Y}$,
which are the groups of Lorentz transformations and
spacetime translations, respectively.
The translation group $\textbf{Y}$ forms a spacetime lattice.
Each vector of $\textbf{Y}$ is a linear combination of its primitive vectors.
$n_0$, $n_1$, $n_2$ and $n_3$ are integers.
$r_1$, $r_2$ or $r_3$ are the primitive vectors of
the corresponding Bravais lattice, while $r_0$ is the temporal primitive vector.
In the expression of $r_0$ and $r_1$, $a_1$ denotes the lattice constant,
$c$ denotes the speed of light, and $g \geq 2$ is an integer generator which
determines the shape of $\textbf{Y}$. The discrete Lorentz group $\textbf{L}$
contains the elements of $\textbf{R}$, as well as the Lorentz boosts $B_{\vec{v}_{j}}$.
The velocity $\vec{v}_{j}$ in the Lorentz boost must be in the direction of $r_1$,
and can only take some discrete values ($j$ is an integer).}
\end{table}
The paper is organized as follows. In Sec.~\ref{sec:def}, we review
the discrete Poincar\'{e} group in 1+1 dimensions, and then introduce the
method of generalization to higher dimensions. Sec.~\ref{sec:2d} and~\ref{sec:3d}
contribute to 1+2 and 1+3 dimensions, respectively. Sec.~\ref{sec:dis} discusses
the possible way of observing discrete Poincar\'{e} symmetry in crystals.

\section{Discrete Poincar\'{e} symmetry}
\label{sec:def}

Crystals have discrete STS. Their local properties change periodically
in space. In 1+1 dimensions, a crystal
looks the same under a spatial translation of coordinates if and only if
the translation distance is an integer times of the lattice constant.
Between two reference frames that are moving relative to each other, the coordinates
transform as a Lorentz transformation.
Usually, crystals look different in different reference frames.
But Ref.~[\onlinecite{Wang}] showed
that it is possible for crystals to look the same if the relative velocity takes
some specific values. The cost is that the properties of crystals must
change with time at some specific periods. This kind of symmetry is called
the discrete Poincar\'{e} symmetry.

Ref.~[\onlinecite{Wang}] discussed in general how to construct
a periodic function of spacetime that is invariant under a group of Lorentz transformations.
STS of a crystal is determined by its Bravais lattice. The 1D Bravais
lattice can be extended in the time direction to form a 1+1-dimensional
spacetime lattice. For simplicity, let us view a crystal as a spacetime lattice.
Under specific Lorentz transformations, this lattice keeps
the same in spite of length contraction and time dilation. This is possible
if the spatial direction after transformation is still in
one of the lattice directions. The spatial directions vary from
one reference frame to the other, but the lattice constant, i.e. the distance between
two neighbor sites in the spatial direction, keeps the same. Notice that two
spatial neighbor sites in one reference frame are not spatial neighbors
in the other because they are not simultaneous any more.

All the coordinate transformations that keep a crystal invariant make up a group, dubbed
the discrete Poincar\'{e} group. It contains the discrete Lorentz transformations
and the spacetime translation of lattice vectors.
The element of a discrete Poincar\'{e} group is denoted by
$\Lambda(L, r)$, which means a Lorentz transformation $L$ followed by a spacetime
translation $r$. In 1+1 dimensions, $L$ is a 2-by-2 matrix and $r$ is a two-components vector.
According to definition, the product of two group elements reads
\begin{equation}\label{eq:mulrule}
\Lambda(L',r')\Lambda(L,r)
=\Lambda(L'L, L' r +r').
\end{equation}
The discrete Poincar\'{e} group is found to be
\begin{equation}\label{eq:P1d}
\mathcal{P}=\left\{ \Lambda\left( L_{v_j}, r_{n_0 n_1}\right) \bigg| j,n_0,n_1=0,\pm 1, \pm 2, \cdots \right\},
\end{equation}
where $j,n_0$ and $n_1$ are arbitrary integers.
In the Lorentz transformation,
$v_j$ denotes the velocity of one reference frame relative to the other,
taking the value $v_j/c=\textbf{sign}(j) \sqrt{1-4/m_j^2}$
with $c$ the speed of light.
And $m_j$ is an integer sequence satisfying $m_{j+1}=gm_j-m_{j-1}$,
where $g$ is the generator of the sequence, $m_0=2$, $m_1=g$ and $m_{-j}=m_j$.
The translation vector $r_{n_0 n_1}$ reads
\begin{equation}\label{eq:lattice}
\begin{split}
r_{n_0 n_1}= n_0 r_0 + n_1 r_1,
\end{split}
\end{equation}
where $r_1=\left(0,a_1\right)^T$ with $a_1$
denoting the lattice constant. And $r_0$ equals $\left( a_1 \sqrt{g^2-4}/{(2c)}, 0 \right)^T$ for even $g$ or
$\left( a_1 \sqrt{g^2-4}/{(2c)}, a_1/2 \right)^T$ for odd $g$.
Note that the first component of a vector denotes the time coordinate, while the second denotes the
space coordinate.

A discrete Poincar\'{e} group is determined by its generator $g$ ($g\geq 2$).
As $g=2$, we have $m_j \equiv 2$ and $v_j \equiv 0$. $\mathcal{P}$ contains
no Lorentz transformation, and the translation vector $r_{n_0 n_1}$ is purely spatial.
This is the symmetry group of 1D crystals in the orthodox view.

In the case $g>2$, the spatial translations ($n_0=0$) in $\mathcal{P}$ keep the same.
But $\mathcal{P}$ contains additional discrete Lorentz transformations
and discrete TTS.
As $g>2$ is even, the points $r_{n_0 n_1}$ form a rectangular
lattice in 1+1-dimensional spacetime. This lattice gives the spacetime
translational symmetry. Both TTS and STS are discrete, which are connected
to each other by the discrete Lorentz rotations $L_{v_j}$.
The period of TTS is $\sqrt{g^2-4}a/(2c)$.
As $g$ is odd, the points $r_{n_0n_1}$ form a centered rectangular lattice.
The period of TTS becomes $\sqrt{g^2-4}a/c$.
The velocity $v_j$ in the Lorentz rotation can only take discrete
values. For example, $v_j$ takes $0, \sqrt{5}c/3, 3\sqrt{5}c/7, \cdots $ at $g=3$,
or $0, \sqrt{3}c/2, 4\sqrt{3}c/7,\cdots$ at $g=4$.

$\mathcal{P}$ is the subgroup of the continuous Poincar\'{e} group
for whatever $g$. It is reasonable to guess
that $\mathcal{P}$ at $g>2$ is also the symmetry group of some crystals.
If a crystal chooses $\mathcal{P}$ at $g>2$ as its symmetry group,
it looks the same after a coordinate translation only if the translation vector is $r_{n_0 n_1}$.
It means that the local properties are varying not only with space but also with time.
They are periodic functions of spacetime coordinates.

A few more words are necessary for explaining the physical meaning of $v_j$.
Let us view a crystal as a chain of atoms with the distance between
two neighbors being $a$. The discrete TTS requires that the atoms are
moving periodically even in the rest frame (like a lattice vibration).
If all the atoms are oscillating in the same phase,
$a$ becomes the period in the spatial direction (the lattice constant).
But in a moving reference frame, the atoms are not oscillating in the same phase anymore,
because simultaneity depends on the reference frame. Starting from an atom $A$,
we will find that the movement of its neighbor is now behind $A$.
But since the oscillation is periodic, after a few atoms,
we may again find an atom $B$ that is oscillating in the same phase as $A$.
Now the lattice constant becomes the distance between $A$ and $B$.
But due to the length contraction, this distance is indeed $a$ in the moving
reference frame. For this to happen, the contraction must be strong, since the distance between
$A$ and $B$ in the rest frame is a few times of $a$. This explains why
$v_j$ is comparable to the speed of light.

Eq.~(\ref{eq:P1d}) gives the discrete Poincar\'{e} groups in 1+1 dimensions.
In this paper, we generalize to 1+2 and
1+3 dimensions. For this purpose, we define the symmorphic Poincar\'{e} group.
The element $\Lambda(L,r)$ is a pure Lorentz transformation as $r=0$,
or a pure translation as $L=1$ (the identity matrix). We use $\textbf{L}=\left\{L\right\}$
to denote a group of Lorentz transformations,
and $\textbf{Y}=\left\{r \right\}$ to denote a group of translations.
If the set $\mathcal{P}$ is a direct product of $\textbf{L}$ and $\textbf{Y}$ and
$\mathcal{P}$ is a group under the multiplication rule~(\ref{eq:mulrule}),
we say that $\mathcal{P}$ is a symmorphic Poincar\'{e} group.
This definition is similar to the symmorphic space group in crystallography.
In 1+1 dimensions, the discrete Poincar\'{e} group is indeed a symmorphic group.

In $d \geq 2$ spatial dimensions, the Lorentz transformation $L\in \textbf{L}$
is a (1+d)-by-(1+d) matrix. The translation $r\in \textbf{Y}$ is a $(1+d)$-dimensional vector.
$L$ acting on $r$ gives the transformation of a vector between different reference frames.
If $Lr$ and $L^{-1}r$ are both the elements
of $\textbf{Y}$ for each $r\in \textbf{Y}$, we say that $\textbf{Y}$
is invariant under $L$. Note that $\textbf{Y}$ is indeed a lattice in $(1+d)$-dimensional spacetime.
$\textbf{Y}$ being invariant under $L$ means that this lattice keeps the same after
the spacetime rotation $L$.
If $\textbf{Y}$ is invariant under each $L \in \textbf{L}$, we
say that $\textbf{Y}$ is invariant under $\textbf{L}$. We construct
discrete Poincar\'{e} groups based on next fact: $\textbf{L} \times \textbf{Y}$
is a symmorphic Poincar\'{e} group if and only if $\textbf{Y}$ is invariant under $\textbf{L}$.
The proof is given in appendix~\ref{sec:app1}.

In $d\geq 2$ spatial dimensions, symmetry operations include rotation, reflection,
inversion and improper reflection. These operations are denoted by $R$.
And we use $B$ to denote a Lorentz boost
(a symmetric matrix in the unit $c=1$). In general, a Lorentz transformation
in 1+d dimensions can be expressed as $L=RB$.
In $d\geq 2$ spatial dimensions, the group $\textbf{L}$ contains not only
Lorentz boosts, but also spatial rotations, reflections, etc..
Indeed, it is not enough to identify a reference frame by just giving its velocity.
Because different reference frames may differ by a rotation.
This complexity causes all the difficulties in the construction of
discrete Poincar\'{e} groups.

In our approach, we start from a d-dimensional Bravais lattice,
extending it in the time direction to form a (1+d)-dimensional lattice $\textbf{Y}$.
Therefore, $\textbf{Y}$ contains the Bravais lattice as its part.
This is what we require for $\mathcal{P}=\textbf{L}\times \textbf{Y}$
being the symmetry group of a Bravias lattice.

We then check if $\textbf{Y}$ is invariant under the Lorentz boosts,
rotations, reflections, etc.. These operations make up the group $\textbf{L}$.
$\textbf{L}$ contains the point group of the Bravais lattice.
For example, for the 2D square lattice, $\textbf{L}$
must contain the spatial rotations of angles $0$, $\pi/2$, $\pi$ or $3\pi/2$.
If $\textbf{L}$ contains nothing more than
the point group, $\textbf{Y}$ is obviously invariant under $\textbf{L}$
(a Bravais lattice is invariant under its point group according to definition).
In this case, we obtain a trivial symmorphic Poincar\'{e} group
with no Lorentz transformation. It is
the symmetry group of crystals in the orthodox view.

For $\mathcal{P}$ to be nontrivial, $\textbf{L}$
must contain at least one Lorentz boost $B_{\vec{v}}$ with
$\vec{v}\neq 0$~\bibnote{There exist possibilities that $\textbf{L}$ contains no Lorentz boost
but only composite Lorentz transformations ($L=RB$), which are not discussed in
this paper.}. We find that
$\textbf{Y}$ is invariant under $B_{\vec{v}}\in \textbf{L}$ only if $\vec{v}$ is in the
lattice direction, that is $\vec{v}$ connects at least two sites of the Bravais lattice
(see appendix~\ref{sec:app2} for the proof). Without loss of generality, we suppose that $\vec{v}$
is in the $x$-direction. The sites of $\textbf{Y}$ in the $x$-axis
then form a 1D Bravais lattice. Eq.~(\ref{eq:lattice}) has told us the unique way of
extending a 1D Bravais lattice into a (1+1)-dimensional spacetime lattice.
In this way, we naturally obtain the sublattice of $\textbf{Y}$ in the $t$-$x$ plane.
We can then obtain the whole $\textbf{Y}$ and $\mathcal{P}=\textbf{L}\times \textbf{Y}$
(see the detailed derivation in appendix~\ref{sec:app3} and~\ref{sec:app4}).
$\mathcal{P}$ is the subgroup of the continuous Poincar\'{e} group,
at the same time contains the space group of Bravais lattice as its subgroup.
In next sections, we enumerate $\mathcal{P}$ in 1+2 and 1+3 dimensions.

\section{Two dimensional Bravais lattices}
\label{sec:2d}

In two spatial dimensions, there are five Bravais lattices: oblique, rectangular, centered rectangular, hexagonal
and square lattices. There exist symmorphic Poincar\'{e} groups on the rectangular lattice,
but no symmorphic Poincar\'{e} groups on the other four lattices (see table~\ref{tab:1}).
The derivation is given in appendix~\ref{sec:app3}.

Recall that an element of the symmorphic Poincar\'{e} group
is denoted as $\Lambda(L,r) $, where $L$ is an element of $\textbf{L}$
and $r$ is an element of $\textbf{Y}$. Here $\textbf{L}$ is the group of Lorentz
transformations and $\textbf{Y}$ is the group of translations.
$r$ is a vector in 1+2-dimensional spacetime, and can be generally
expressed as $r=(t,x,y)^T$ where $t$ denotes the time and $x$ and $y$ denote the space coordinates. 
Two primitive vectors of the rectangular lattice are $r_1=(0,a_1,0)^T$
and $r_2=(0,0,a_2)^T$ with $a_1\neq a_2$.

\begin{figure}
\centering
\includegraphics[width = 0.8 \linewidth]{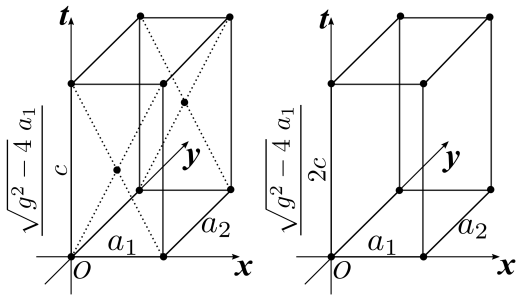}
\caption{A unit cell of $\textbf{Y}$ for 2D rectangular lattice. The
left panel is for odd $g$, while the right panel is for even $g$.}\label{fig:1}
\end{figure}

The symmorphic Poincar\'{e} group
is determined by an integer generator $g\geq 2$.
All the Lorentz boosts in $\textbf{L}$ must be in the same direction.
It is either in the $x$-direction or in the $y$-direction.
Without loss of generality, we suppose it to be in the $x$-direction.
The temporal primitive vector of $\textbf{Y}$ is then
$r_0^{(e)}=(\sqrt{g^2-4} \ a_1/{(2c)}, 0,0)^T$ for even $g$, or
$r_0^{(o)}=(\sqrt{g^2-4}\ a_1/{(2c)}, a_1/2,0)^T$ for odd $g$.
An arbitrary vector of $\textbf{Y}$ can be expressed as
\begin{equation}
r_{n_0 n_1n_2}= n_0 r_0 + n_1 r_1 +n_2r_2,
\end{equation}
where $n_0$, $n_1$ and $n_2$ are integers.
Fig.~\ref{fig:1} shows a unit cell of $\textbf{Y}$.
$\textbf{Y}$ in 1+2 dimensions is indeed an extension of
the 1+1-dimensional spacetime lattice along a perpendicular direction.
$\textbf{Y}$ is an orthorhombic lattice for even $g$ or a base-centered
orthorhombic lattice for odd $g$.

The Lorentz boost in $\textbf{L}$ can be expressed as
\begin{equation}
\begin{split}
B_{{v}_j} = \left( \begin{array}{ccc} 
\displaystyle\frac{1}{\sqrt{1-v_j^2/c^2}} & \displaystyle\frac{-v_j/c^2}{\sqrt{1-v_j^2/c^2}} & 0 \\
\displaystyle\frac{-v_j}{\sqrt{1-v_j^2/c^2}} & \displaystyle\frac{1}{\sqrt{1-v_j^2/c^2}} & 0 \\
0 & 0 & 1 \end{array}\right),
\end{split}
\end{equation}
where $v_j=\textbf{sign}(j)\sqrt{1-4/m^2_j(g)} \ c$ takes the same value as
in 1+1 dimensions. $B_{{v}_j}$ acting on a vector keeps its $y$-component invariant.
The transformation is purely within the $t$-$x$ plane.

We use $\textbf{R}$ to denote the point group of a Bravais lattice.
For the rectangular lattice, $\textbf{R}$ contains a rotation of angle $\pi$ in the
$x$-$y$ plane, a reflection across the $x$-axis and a reflection across
the $y$-axis. The discrete Lorentz group $\textbf{L}$ can then be expressed as
\begin{equation}
\textbf{L} = \left\{ R B_{v_j} \big| R\in \textbf{R} \ and \ j= 0, \pm1, \pm2, \cdots \right\}.
\end{equation}
The element of $\textbf{L}$ is the product of a spatial operation and a Lorentz boost.
In the case $g=2$, $B_{v_j}$ is the identity matrix and $\textbf{L}$ reduces to $\textbf{R}$,
that is the Lorentz symmetry is absent. For $g>2$, $\textbf{L}$ gives a discrete Lorentz
symmetry in the $x$-direction, but no Lorentz symmetry in the $y$-direction.

\section{Three dimensional Bravais lattices}
\label{sec:3d}
   
\begin{figure}
\centering
\includegraphics[width = 0.8 \linewidth]{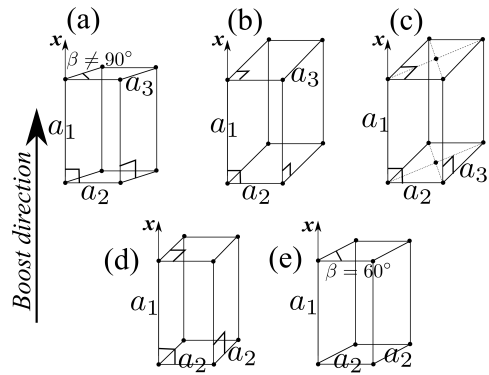}
\caption{The boost direction on the (a) monoclinic, (b) orthorhombic,
(c) base-centered orthorhombic, (d) tetragonal and (e) hexagonal lattices.
$a_1$, $a_2$ and $a_3$ denote the lattice constants. And we use $\beta$
to denote the angle between two primitive vectors in the $y$-$z$ plane when
it is not $90^{\circ}$.}
\label{fig:2}
\end{figure}

In three spatial dimensions, there are 14 Bravais lattices. There
exists symmorphic Poincar\'{e} symmetry on the monoclinic,
orthorhombic, base-centered orthorhombic,
tetragonal and hexagonal lattices (see table~\ref{tab:1}).
The derivation is given in appendix~\ref{sec:app4}.

We denote a vector in 1+3-dimensional spacetime as $(t,x,y,z)^T$.
Again, $\mathcal{P}=\textbf{L}\times \textbf{Y}$ denotes the symmorphic Poincar\'{e}
group. The Lorentz boosts in $\textbf{L}$ are supposed to be in the
$x$-direction without loss of generality.
Similar to 1+2-dimensional case, the Lorentz boost is expressed as
\begin{equation}
\begin{split}
B_{{v}_j} = \left( \begin{array}{cccc} 
\displaystyle\frac{1}{\sqrt{1-v_j^2/c^2}} & \displaystyle\frac{-v_j/c^2}{\sqrt{1-v_j^2/c^2}} & 0 & 0 \\
\displaystyle\frac{-v_j}{\sqrt{1-v_j^2/c^2}} & \displaystyle\frac{1}{\sqrt{1-v_j^2/c^2}} & 0 & 0 \\
0 & 0 & 1 & 0 \\
0 & 0 & 0 & 1 \end{array}\right),
\end{split}
\end{equation}
where $v_j$ takes the same value as in 1+1 or 1+2 dimensions.
And $\textbf{L}$ is again $\left\{ R B_{v_j} \big| R\in \textbf{R} \ and \ j= 0, \pm1, \pm2, \cdots \right\}$,
where $\textbf{R}$ is the point group of the corresponding Bravais lattice.
There is no Lorentz symmetry in the $y$-$z$ plane.

$\textbf{Y}$ in 1+3 dimensions has four primitive vectors. We use
$r_1$, $r_2$ and $r_3$ to denote the three primitive vectors of the
Bravais lattice. If $r_1$ is chosen to be $(0,a_1,0,0)^T$,
the temporal primitive vector is then
$r_0^{(e)}=(\sqrt{g^2-4} \ a_1/{(2c)}, 0,0,0)^T$ for even $g$, or
$r_0^{(o)}=(\sqrt{g^2-4}\ a_1/{(2c)}, a_1/2,0,0)^T$ for odd $g$.
A vector of $\textbf{Y}$ can be expressed as
\begin{equation}\label{eq:tr}
r_{n_0n_1n_2n_3}= n_0 r_0 + n_1 r_1 + n_2 r_2 + n_3 r_3,
\end{equation}
where $n_0$, $n_1$, $n_2$ and $n_3$ are integers.

Note that $r_1$ cannot be chosen arbitrarily. It is required that
$r_2$ and $r_3$ be both perpendicular to $r_1$.
Fig.~\ref{fig:2} displays the direction of $r_1$ on different Bravais lattices.
The monoclinic, orthorhombic, base-centered orthorhombic,
tetragonal and hexagonal lattices can be viewed as the five 2D Bravais lattices extended
in the perpendicular direction, respectively.
And this perpendicular direction must be chosen to the direction of $r_1$.
Especially, in the tetragonal lattice, $r_1$ must lie in the vertical direction
in which the lattice constant is different from those in the other two directions.

Again, the discrete Poincar\'{e} group $\mathcal{P}$ is uniquely determined
by the integer generator $g\geq 2$. In the case $g=2$,
we obtain $B_{v_j} \equiv 1$ and $\textbf{L}\equiv \textbf{B}$.
And $r_0=0$ indicates a continuous TTS.
The symmetry group reduces to the space group of Bravais lattices.

In the case $g>2$, $v_j(g)=\textbf{sign}(j) \sqrt{1-4/m_j^2} \ c$ is nonzero for $j\neq 0$.
When one reference frame is moving at $v_j$ relative to the other,
the Bravais lattice looks the same for them.
And the temporal primitive vector is now nonzero,
indicating that TTS is broken into a discrete one.

\section{Discussion}
\label{sec:dis}

We have enumerated the symmorphic Poincar\'{e} groups on 2D and 3D
Bravais lattices. An interesting question is whether the symmorphic Poincar\'{e}
symmetry does exist in real crystals. According to our results,
it is possible to find such a symmetry only
in crystals based on the monoclinic, orthorhombic, base-centered orthorhombic,
tetragonal or hexagonal lattices.
For examples, graphite and rutile are based on the
hexagonal and tetragonal Bravais lattices, respectively.

If a crystal has symmorphic Poincar\'{e} symmetry, it keeps invariant
under a coordinate transformation of $\mathcal{P}=\textbf{L}\times \textbf{Y}$,
where $\textbf{L}$ and $\textbf{Y}$ are determined by the generator $g$.
Each crystal has a unique $g$.
To be more precise, when we describe the motion of electrons (or phonons) in this crystal,
the Lagrangian density or equation of motion need to keep invariant under $\mathcal{P}$, which imposes
a strong constraint on the possible form of them. Here, the Lagrangian (or equation of motion)
is an effective one, only for the particles that we are interested in. The full Lagrangian
for the electrons, nuclei and their interactions has of course the continuous Poincar\'{e} symmetry.
One can think that the effective one is derived from the full one
by using some mean-field theory.
Ref.~[\onlinecite{Wang}] showed how to construct
such an effective Lagrangian from the symmetry principle.
One starts from a Lagrangian with continuous Poincar\'{e}
symmetry and then replaces the constants (coupling or mass) by a function $f(r)$
that has the symmetry $\mathcal{P}$. Especially, $f(r)$
is invariant under a translation of vector $r_{n_0n_1n_2n_3}$,
where $r_{n_0n_1n_2n_3}$ is given by Eq.~(\ref{eq:tr}).
As $n_0=0$, this means that $f(r)$ is a periodic function of space.
As $n_0 \neq 0$, we find that $f(r)$ is also a periodic function of time:
\begin{equation}
f(t,x,y,z)=f(t+T,x,y,z).
\end{equation}
It is easy to see $T={\sqrt{g^2-4} \ a_1}/{(2c)}$ for even $g$
or $T={\sqrt{g^2-4} \ a_1}/{c}$ for odd $g$. Here $a_1$ is the lattice constant.
Since the coupling $f$ in the equation of motion is a periodic function,
we expect the solutions to be also periodic functions
with the same period. Therefore, the local properties of the crystal should
change periodically in the spacetime.

If the generator of a crystal is $g=2$, we obtain $T= 0$. In this case,
$f$ is independent of time, so are the local properties.
This is what we usually think of. But there exist the other possibilities.
If the generator of a crystal is $g>2$, its local properties
change periodically with time, even in the absence of external driving.
This is the exclusive feature of symmorphic Poincar\'{e} symmetry.
An experiment searching for the time periodicity
in crystals will then clarify whether there exists symmorphic Poincar\'{e} symmetry
or not. Let us take graphite as an example. Its lattice constant is $a_1 \approx 6.7 \times 10^{-10}m$.
Its time period is then $T  =\sqrt{g^2-4}  \times 1.1 \times 10^{-18} s$ for even $g$
or $T  = \sqrt{g^2-4} \times 2.2  \times 10^{-18} s$ for odd $g$.
The time periods of typical crystals are very small. This may explain why
the symmorphic Poincar\'{e} symmetry has not been observed up to now.
An alternative way of observing the time periodicity would be by using the Floquet effect.
The function $f(r)$ in the equation of motion can be treated as a periodically-driving potential.
In the presence of it, we expect the system to absorb radiation of the frequency $1/T$. Therefore,
a peak at this frequency in the absorption spectrum would also support
the existence of discrete Poincar\'{e} symmetry. We use again graphite as an example.
The frequency is $\left({g^2-4}\right)^{-1/2} \times 9.1 \times 10^{17} \text{Hz}$
for even $g$ or $\left({g^2-4}\right)^{-1/2} \times 4.5 \times 10^{17} \text{Hz}$
for odd $g$. It is in the frequency range of X-rays.

This paper focuses on the classification of discrete Poincar\'{e} symmetry.
In future, we expect to study the effective field theories
with this symmetry and how to obtain them from
a fundamental theory with continuous Poincar\'{e} symmetry.

\section*{Acknowledgement}
This work is supported by NSFC under Grant No.~11774315.
P. Wang is also supported by the Junior Associates program of the Abdus
Salam International Center for Theoretical Physics.

Xiuwen Li, Jiaxue Chai and Huixian Zhu contribute equally to this paper.

\appendix

\section{Proof of $\textbf{L}\times \textbf{Y}$ being a symmorphic Poincar\'{e} group}
\label{sec:app1}

Suppose $\textbf{L}=\left\{ L \right\}$ is a group of Lorentz transformations and
$\textbf{Y}=\left\{ r\right\}$ is a group of translations. We express the set $\textbf{L}\times \textbf{Y}$
as $\mathcal{P}=\left\{ \Lambda(L,r) | L\in \textbf{L}, r\in \textbf{Y}\right\}$.
According to definition, $\mathcal{P}$ is a symmorphic Poincar\'{e} group
if and only if $\mathcal{P}$ is a group under the multiplication rule~(\ref{eq:mulrule}).
Therefore, our destination is to prove that $\mathcal{P}$ is a group
under the rule~(\ref{eq:mulrule}) if and only if $\textbf{Y}$ is invariant
under $\textbf{L}$.

First, we prove that $\mathcal{P}$ is a group if $\textbf{Y}$ is invariant
under $\textbf{L}$. The associativity of the multiplication rule~(\ref{eq:mulrule}) is obvious.
If $\Lambda(L,r)$ and $\Lambda(L',r')$ are two elements of $\mathcal{P}$,
we have $L,L'\in \textbf{L}$ and $r,r'\in \textbf{Y}$ according to definition.
By using Eq.~(\ref{eq:mulrule}), we obtain $\Lambda(L,r)\Lambda(L',r')= \Lambda(LL', Lr'+r)$.
Since $\textbf{L}$ is a group, $LL' \in \textbf{L}$ is obvious.
And because $\textbf{Y}$ is invariant under $\textbf{L}$, we obtain $Lr' \in \textbf{Y}$
and then $ Lr'+r \in \textbf{Y}$ ($ \textbf{Y}$ is a group). Therefore, $\Lambda(LL', Lr'+r)$
must be an element of $\mathcal{P}$.
The closure of $\mathcal{P}$ is proved.
$\textbf{Y}$ and $\textbf{L}$ are groups, so that they contain identity elements.
The identity element of $\textbf{Y}$ is $r=0$ (no translation),
and the identity element of $\textbf{L}$ is $L=1$ (no Lorentz rotation).
By using Eq.~(\ref{eq:mulrule}), it is easy to see $\Lambda(1,0)
\Lambda(L,r) = \Lambda(L,r) \Lambda(1,0)=\Lambda(L,r) $
for arbitrary $L$ and $r$. Therefore, $\Lambda(1,0)$ is the identity element of $\mathcal{P}$.
The existence of identity element is proved.
If $\Lambda(L,r)$ is an element of $\mathcal{P}$, we have $L\in \textbf{L}$
and then $L^{-1} \in \textbf{L}$, and $r\in \textbf{Y}$. Since $\textbf{Y}$ is invariant
under $\textbf{L}$, we obtain $L^{-1}r \in \textbf{Y}$ and then $-L^{-1}r \in \textbf{Y}$.
Therefore, $\Lambda(L^{-1},-L^{-1}r)$ is also an element of $\mathcal{P}$ according to definition.
And it is easy to see that $\Lambda(L^{-1},-L^{-1}r)$ is the inverse of $\Lambda(L,r)$.
The existence of inverse element is proved.
As above, if $\textbf{Y}$ is invariant under $\textbf{L}$, $\mathcal{P}$ must be a group.

Second, we prove that $\mathcal{P}$ is not a group if $\textbf{Y}$ is not invariant
under $\textbf{L}$. If $\textbf{Y}$ is not invariant under $\textbf{L}$, there exist
$L\in \textbf{L}$ and $r\in \textbf{Y}$ so that $Lr \not\in  \textbf{Y}$.
Because $\textbf{L}$ and $\textbf{Y}$ are groups and the set $\mathcal{P}$
is the direct product of $\textbf{L}$ and $\textbf{Y}$, we know that $\Lambda(L,0)$, $\Lambda(1,r)$
and $\Lambda(L^{-1},0)$ are the elements of $\mathcal{P}$.
Their product is $\Lambda(L,0)\Lambda(1,r)\Lambda(L^{-1},0)= \Lambda(1,Lr)$.
But $\Lambda(1,Lr)$ is not an element of $\mathcal{P}$. Therefore, $\mathcal{P}$
is not closed under the rule~(\ref{eq:mulrule}). $\mathcal{P}$ is not a group.

As above, $\mathcal{P}$ is a group if and only if $\textbf{Y}$ is invariant
under $\textbf{L}$.

\section{Proof of $\vec{v}$ being in the lattice direction}
\label{sec:app2}

Suppose $\textbf{Y}$ is a spacetime lattice obtained by extending a Bravais lattice
in the time direction. $B_{\vec{v}} $ is a Lorentz boost, and it is an element of $\textbf{L}$.
In this appendix, we prove that $\textbf{Y}$ is invariant under $\textbf{L}$
only if $\vec{v}$ is in the lattice direction.

In $1+d$ dimensions, we can express the spacetime coordinates as $(t, x_1, \cdots, x_d)^T$.
Without loss of generality, we suppose that $\vec{v}$ is confined to the $x_1$-direction.
The matrix $B_{\vec{v}}$ then looks like
\begin{equation}
\begin{split}
B_{\vec{v}} = \left( \begin{array}{cccc} 
\displaystyle\frac{1}{\sqrt{1-v^2/c^2}} & \displaystyle\frac{-v/c^2}{\sqrt{1-v^2/c^2}} & 0 & \cdots \\
\displaystyle\frac{-v}{\sqrt{1-v^2/c^2}} & \displaystyle\frac{1}{\sqrt{1-v^2/c^2}} & 0 & \cdots \\
0 & 0 & 1 & \cdots \\ 
\cdots & \cdots & \cdots & \cdots \end{array}\right).
\end{split}
\end{equation}
$B_{\vec{v}}$ is a symmetric matrix if we use the unit $c=1$.
The diagonal elements are all $1$ and the off-diagonal elements
are all zero except for the first two lines and columns. Suppose $r\in \textbf{Y}$
is a Bravais lattice vector, hence its temporal component is zero.
We choose $r=(0, x_1, \cdots, x_d)^T$ with $x_1\neq 0$ without loss of generality.
Indeed, there always exist lattice vectors with nonzero component in the $x_1$-direction,
otherwise, it is not a Bravais lattice.

If $\textbf{Y}$ is invariant under $B_{\vec{v}}$, $B_{\vec{v}} r$ must be
an element of $\textbf{Y}$. And $B_{\vec{v}}\in \textbf{L}$ infers $B_{-\vec{v}}\in \textbf{L}$.
Therefore, $B_{-\vec{v}} r$ is an element of $\textbf{Y}$, hence
$r' = B_{\vec{v}}r+B_{-\vec{v}} r -2r$ is an element of $\textbf{Y}$.
Here we have used the properties of $\textbf{Y}$ being a group.
$r'$ reads $\left( 0 , k x_1 , 0, \cdots \right)^T$ with $k=\frac{2}{\sqrt{1-v^2/c^2}} -2 $.
We have $k>0$ for $v\neq 0$, and then $r'\neq 0$.
According to the definition of $\textbf{Y}$, $r'$ is a Bravais lattice vector since
its temporal component is zero. And $r'$ lies in the $x_1$-axis. This means that
the $x_1$-direction is a lattice direction. We then proved that $\vec{v}$ is in the lattice direction.

\section{(1+2)-dimensional symmorphic Poincar\'{e} groups}
\label{sec:app3}

In this section, we construct the symmorphic Poincar\'{e} groups in 1+2 dimensions.
We use $(t, x, y)^T$ to denote the coordinates in 1+2 dimensions.
Suppose $\mathcal{P}=\textbf{L}\times \textbf{Y}$ is a symmorphic Poincar\'{e} group.
$\textbf{Y}$ is invariant under $\textbf{L}$. For nontrivial $\mathcal{P}$,
$\textbf{L}$ contains at least one Lorentz boost $B_{\vec{v}}$.
And $\textbf{Y}$ is invariant under $B_{\vec{v}}$. As proved in appendix~\ref{sec:app2},
${\vec{v}}$ must be in the lattice direction. Without loss of generality,
we suppose that ${\vec{v}}$ is in the $x$-direction and the lattice constant
in this direction is $a_1$. Therefore, $r_1=(0,a_1,0)^T$ is a primitive vector of
the Bravais lattice, and a primitive vector of $\textbf{Y}$ as well.

$B_{\vec{v}}$ can be expressed as
\begin{equation}
\begin{split}
B_{\vec{v}} = \left( \begin{array}{ccc} 
\displaystyle\frac{1}{\sqrt{1-v^2/c^2}} & \displaystyle\frac{-v/c^2}{\sqrt{1-v^2/c^2}} & 0 \\
\displaystyle\frac{-v}{\sqrt{1-v^2/c^2}} & \displaystyle\frac{1}{\sqrt{1-v^2/c^2}} & 0 \\
0 & 0 & 1 \end{array}\right),
\end{split}
\end{equation}
which is a symmetric matrix in the unit $c=1$.
$B_{\vec{v}}$ rotates $r_1$ into the $t$-$x$ plane. The sublattice of $\textbf{Y}$
in the $t$-$x$ plane is a (1+1)-dimensional spacetime lattice, which
is invariant under $B_{\vec{v}}$. Obviously, this sublattice and $B_{\vec{v}}$
are the elements of a (1+1)-dimensional discrete Poincar\'{e} group.
As reviewed in Sec.~\ref{sec:def}, such a group is determined by
an integer generator $g\geq 2$. For an even $g$, the primitive vectors
of the sublattice are $r_1$ and $r_0^{(e)}=(\sqrt{g^2-4} \ a_1/{(2c)}, 0,0)^T$.
For an odd $g$, the primitive vectors are $r_1$ and $r_0^{(o)}=(\sqrt{g^2-4}\ a_1/{(2c)}, a_1/2,0)^T$.
And the velocity in the Lorentz boost can only take the values
\begin{equation}\label{appeq:vj}
v_j=\textbf{sign}(j)\sqrt{1-4/m_j^2(g)} \ c 
\end{equation}
with $j=0, \pm 1, \pm2 ,\cdots $.

The lattice $\textbf{Y}$ has three primitive vectors in 1+2 dimensions.
Among them, $r_0$ and $r_1$ lie in the $t$-$x$ plane.
While the Bravais lattice has two primitive vectors in the spatial dimensions,
namely $r_1$ and $r_2$.
Obviously, $r_0$, $r_1$ and $r_2$ must be the three primitive vectors of $\textbf{Y}$.

Without loss of generality, we express $r_2$ as $ (0, r_{2x}, r_{2y})^T$.
An arbitrary vector of $\textbf{Y}$ (dubbed a lattice vector)
can be expressed as $n_0 r_0 + n_1 r_1 +n_2 r_2$
with $n_0$, $n_1$ and $n_2$ being integers. $\textbf{Y}$ is invariant under $B_{\vec{v}}$
if and only if $B_{\vec{v}} r_0$, $B_{\vec{v}} r_1$ and $B_{\vec{v}} r_2$ are
lattice vectors. $B_{\vec{v}} r_0$ and $B_{\vec{v}} r_1$ are obviously lattice vectors,
since $r_0$ and $r_1$ are the primitive vectors of the (1+1)-dimensional
sublattice which is invariant under $B_{\vec{v}}$.

Let us study the condition of $B_{\vec{v}} r_2$ being a lattice vector.
The value of $\vec{v} $ is given by Eq.~(\ref{appeq:vj}).
And we already know from Ref.~[\onlinecite{Wang}]
that $B_{\vec{v}_j}=\left(B_{\vec{v}_1}\right)^j$, where the superscript denotes the exponent.
Therefore, $B_{\vec{v}_j} r_2$ is a lattice vector if and only if
$B_{\vec{v}_1} r_2$ and $B_{-\vec{v}_1} r_2$ are lattice vectors.
Note $\left(B_{\vec{v}_1}\right)^{-1}= B_{-\vec{v}_1}$.
And the elementary Lorentz boost is
\begin{equation}
\begin{split}
B_{\pm\vec{v}_1} = \left( \begin{array}{ccc} 
\frac{g}{2} &  \mp \frac{1}{c} \sqrt{\frac{g^2}{4}-1} & 0 \\
\mp \sqrt{\frac{g^2}{4}-1} \ {c}  & \frac{g}{2} & 0 \\
0 & 0 & 1 \end{array}\right).
\end{split}
\end{equation}
Now we obtain two equations:
\begin{equation}
\begin{split}
B_{\vec{v}_1} r_2 = & n_0 r_0 + n_1 r_1 +n_2 r_2 \\
B_{-\vec{v}_1} r_2 = & \bar n_0 r_0 + \bar n_1 r_1 + \bar n_2 r_2 .
\end{split}
\end{equation}
We need to find $r_{2x}$ and $r_{2y}$ so that
$n_0$, $n_1$, $n_2$, $\bar n_0$, $\bar n_1$, $\bar n_2$ are all integers.
It is easy to see that $r_{2x}$ must be an integer times of $a_1$ but $r_{2y}$
can be arbitrary real number. Without loss of generality, we choose $r_{2x}=0$
and rename $r_{2y}$ as $a_2$. The second primitive vector of
the Bravais lattice is then $r_2= (0,0,a_2)^T$.

In one word, if $\mathcal{P}=\textbf{L}\times \textbf{Y}$ is a symmorphic Poincar\'{e}
group and $\textbf{L}$ contains at least one Lorentz boost, two primitive vectors of the Bravais lattice
must be perpendicular to each other and the
velocity of the boost is in the direction of one primitive vector.

With this in mind, we rule out the possibility of oblique, centered rectangular
or hexagonal lattices owning nontrivial symmorphic Poincar\'{e} symmetry.
Further analysis rules out the possibility of the square lattice.
This can be proved by contradiction.
We suppose that a square lattice has the primitive vectors $r_1=(0,a,0)^T$ and $r_2=(0,0,a)^T$.
Its spacetime lattice $\textbf{Y}$ is invariant under the Lorentz boost $B_{\vec{v}}$
with $\vec{v}$ lying in the $x$-direction. $B_{\vec{v}}$ is an element of $\textbf{L}$.
Do not forget that $\textbf{L}$ contains the point group of the square lattice.
In particular, a rotation of angle $\pi/2$ in the $x$-$y$ plane is an element of $\textbf{L}$, which can be
expressed as
\begin{equation}
\begin{split}
R_{\pi/2} = \left( \begin{array}{ccc} 
1 &  0 & 0 \\
0 & 0 & -1 \\
0 & 1 & 0 \end{array}\right).
\end{split}
\end{equation}
$B'=R_{\pi/2} B_{\vec{v}} R_{\pi/2}^{-1}$ is then a Lorentz boost in the $y$-direction.
And $\left(B_{\vec{v}}B'B'B_{\vec{v}}\right) \in \textbf{L}$ is a Lorentz boost of velocity $\vec{v}'$.
$\vec{v}'$ makes an angle $0<\theta<\pi/2$ with the positive $x$-axis.
As proved in above, $\vec{v}'$ must be in a lattice direction of the square lattice,
since $\textbf{Y}$ is also invariant under $\left(B_{\vec{v}}B'B'B_{\vec{v}}\right)$.
And if we use $r'_1$ to denote the primitive vector of the square lattice in the $\vec{v}'$-direction, the other
primitive vector $r'_2$ has to be perpendicular to $r'_1$. But this is impossible
on a square lattice. Therefore, a square lattice cannot have
nontrivial symmorphic Poincar\'{e} symmetry.

In two spatial dimensions, there exist nontrivial symmorphic Poincar\'{e} symmetry only
on the rectangular lattice. If we choose two perpendicular primitive vectors -
$r_1=(0,a_1,0)^T$ and $r_2= (0,0,a_2)^T$ with $a_1\neq a_2$,
the Lorentz boost must be in the direction of $r_1$ or $r_2$.

\section{(1+3)-dimensional symmorphic Poincar\'{e} groups}
\label{sec:app4}

In this section, we construct the symmorphic Poincar\'{e} groups in 1+3 dimensions.
We use $(t, x, y, z)^T$ to denote the coordinates in 1+3 dimensions. Again,
we suppose $\mathcal{P}=\textbf{L}\times \textbf{Y}$ is a symmorphic Poincar\'{e} group.
And $\textbf{L}$ contains at least one Lorentz boost $B_{\vec{v}}$.
${\vec{v}}$ is in the $x$-direction and the lattice constant
in this direction is $a_1$. $r_1=(0,a_1,0, 0)^T$ is a primitive vector of the Bravais lattice.
The other two primitive vectors are supposed to be $r_2=(0,r_{2x},r_{2y}, r_{2z})^T$
and $r_3=(0,r_{3x},r_{3y}, r_{3z})^T$.

Similarly, we can prove that $r_{2x}=0$ and $r_{3x}=0$. $r_2$ and $r_3$
must be perpendicular to $r_1$.
And the temporal primitive vector
of $\textbf{Y}$ is $r_0^{(e)}=(\sqrt{g^2-4} \ a_1/{(2c)}, 0,0, 0)^T$ for even $g$
or $r_0^{(o)}=(\sqrt{g^2-4}\ a_1/{(2c)}, a_1/2,0,0)^T$ for odd $g$.

In three-dimensional Bravais lattices, the boost must be in the direction of one primitive vector
and the other two primitive vectors are in the perpendicular direction.
There are 14 Bravais lattices. Only in the monoclinic, orthorhombic, base-centered orthorhombic,
tetragonal, hexagonal or cubic lattices, there exist two primitive vectors that are
both perpendicular to the third one. But similar to the square lattice in two dimensions,
the cubic lattice cannot have nontrivial symmorphic Poincar\'{e} symmetry.

Notice that the monoclinic, orthorhombic, base-centered orthorhombic,
tetragonal and hexagonal lattices can be created by extending five
2D Bravais lattices in the perpendicular direction. And this direction is indeed the direction
of the Lorentz boost.

\bibliographystyle{apsrev}
\bibliography{literature}

\end{document}